\documentclass[12pt]{article}

\usepackage{a4}
\usepackage{cite, mcite}
\usepackage{mcite}
\usepackage{epsfig}
\usepackage{amsmath}
\usepackage{amssymb}
\usepackage[english]{babel}
\usepackage{fancybox}
\usepackage{color}
\usepackage{hhline}
\usepackage{float}
\usepackage{rotating}
\usepackage[hang,bf]{caption2}
\usepackage{xspace}

\newcommand{\MRM}[1]{\mathrm{#1}}
\newcommand{\ee}{\ensuremath{\mbox{$\MRM{e}^{+}\MRM{e}^{-}$}}\xspace}
\newcommand{\DM}[2]{\ensuremath{\Delta M^{#1}_{#2}}}
\newcommand{\CHI}[2]{\ensuremath{\tilde{\chi}^{#1}_{#2}}}
\newcommand{\M}[1]{\ensuremath{M_{#1}}}
\newcommand{\MCHI}[2]{\M{\CHI{#1}{#2}}}
\newcommand{\W}[1]{\ensuremath{\MRM{W^{#1}}}}
\newcommand{\Z}[1]{\ensuremath{\MRM{Z^{#1}}}\xspace}
\newcommand{\GeV}{\ensuremath{\MRM{GeV}}\xspace}
\newcommand{\mrad}{\ensuremath{\MRM{mrad}}\xspace}
\newcommand{\ETg}{\ensuremath{E_\MRM{T}^\gamma}\xspace}
\newcommand{\SQRTS}{\ensuremath{\sqrt{s}}}
\newcommand{\eeZ}[0]{\mathrm{e}^+\mathrm{e}^-\to{(\mathrm{Z}^0/\gamma)}^\star\to}
\newcommand{\Evis}[0]{\ensuremath{E_\mathrm{vis}}\xspace}
\newcommand{\CL}[0]{\ensuremath{\mathrm{C.L.}}\xspace}
\newcommand{\ud}[0]{\ensuremath{\mathrm{d}}}
\newcommand{\RB}[1]{\raisebox{1.5ex}[-1.5ex]{#1}}

\newcommand{\qqp}{\mbox{$\MRM{q\overline{q}^\prime}$}}

\begin{document}
\begin{titlepage}    
\begin{center}{\large   EUROPEAN ORGANIZATION FOR NUCLEAR RESEARCH    
}\end{center}\bigskip    
\begin{flushright}    
       CERN-EP/2002-063   \\ 25 July 2002
\end{flushright}
\bigskip\bigskip\bigskip\bigskip\bigskip
\begin{center}{\huge\bf  Search for Nearly Mass-Degenerate Charginos
    and Neutralinos at LEP  
}\end{center}\bigskip\bigskip
\begin{center}{\LARGE The OPAL Collaboration
}\end{center}\bigskip\bigskip
\bigskip\begin{center}{\large  Abstract}\end{center}
A search was performed for charginos with masses close to the mass of
the lightest neutralino in \ee collisions at centre-of-mass energies
of 189--209 \GeV recorded by the OPAL detector at LEP. Events were
selected if they had an observed high-energy photon from initial state
radiation, reducing the dominant background from two-photon scattering
to a negligible level. No significant excess over Standard Model
expectations has been observed in the analysed data set corresponding
to an integrated luminosity of 570 pb$^{-1}$. Upper limits were
derived on the chargino pair-production cross-section, and lower
limits on the chargino mass were derived in the context of the Minimal
Supersymmetric Extension of the Standard Model for the gravity and
anomaly mediated Supersymmetry breaking scenarios. 

\bigskip\bigskip
\begin{center}{\large
(To be submitted to Eur. Phys. J.)
}\end{center}

\end{titlepage}
\begin{center}{\Large        The OPAL Collaboration
}\end{center}\bigskip

\begin{center}{
G.\thinspace Abbiendi$^{  2}$,
C.\thinspace Ainsley$^{  5}$,
P.F.\thinspace {\AA}kesson$^{  3}$,
G.\thinspace Alexander$^{ 22}$,
J.\thinspace Allison$^{ 16}$,
P.\thinspace Amaral$^{  9}$, 
G.\thinspace Anagnostou$^{  1}$,
K.J.\thinspace Anderson$^{  9}$,
S.\thinspace Arcelli$^{  2}$,
S.\thinspace Asai$^{ 23}$
% ...
}\end{center}\bigskip
\bigskip
$^{  1}$School of Physics and Astronomy, University of Birmingham,
Birmingham B15 2TT, UK
\newline
$^{  2}$Dipartimento di Fisica dell' Universit\`a di Bologna and INFN,
I-40126 Bologna, Italy
\newline
$^{  3}$Physikalisches Institut, Universit\"at Bonn,
D-53115 Bonn, Germany
\newline
% ...
$^{  5}$Cavendish Laboratory, Cambridge CB3 0HE, UK
\newline
% ...
$^{  9}$Enrico Fermi Institute and Department of Physics,
University of Chicago, Chicago IL 60637, USA
\newline
% ...
$^{ 16}$Department of Physics, Schuster Laboratory, The University,
Manchester M13 9PL, UK
\newline
% ...
$^{ 22}$Department of Physics and Astronomy, Tel Aviv University,
Tel Aviv 69978, Israel
\newline
$^{ 23}$International Centre for Elementary Particle Physics and
Department of Physics, University of Tokyo, Tokyo 113-0033, and
Kobe University, Kobe 657-8501, Japan
\newline
% ...

\section{Introduction}
\label{sec:introduction}
Supersymmetric (SUSY) extensions~\cite{Golfand:1971iw, *Volkov:1973ix,
  *Wess:1974tw} of the Standard Model (SM) provide a promising
approach to overcoming shortcomings of the SM. By introducing
supersymmetric partners for each SM particle, differing in spin by 1/2
unit from the SM particles, the hierarchy problem can be solved and
large radiative corrections to the Higgs mass cancel out. The simplest
potentially realistic supersymmetric field theory is the Minimal
Supersymmetric Extension of the Standard Model (MSSM), which
introduces only a minimum set of additional particles. Among these new
particles are the gauginos and higgsinos which are the fermionic
partners of the gauge and Higgs bosons, respectively. The charginos
(\CHI{\pm}{i=1,2}) are the mass eigenstates formed by the mixing of
the fields of the charged gauginos and higgsinos. The neutralinos
(\CHI{0}{j=1,\dots,4}) are correspondingly formed by the mixing of the
neutral gauginos and higgsinos.

In this paper the hypothesis of $R$-parity conservation is made.
$R$-parity conserving SUSY scenarios allow only the pair-production of
SUSY particles. In addition, there must exist a lightest, neutral,
weakly interacting SUSY particle (LSP) which terminates the decay
chain of any SUSY particle. In the following it is assumed that the
lightest neutralino is the LSP. The lack of experimental evidence for
supersymmetric particles leads to the assumption that SUSY particles
are much heavier than their SM partners. Therefore, SUSY cannot be an
exact symmetry of nature. If it exists it must be a broken symmetry.
Further assumptions and constraints are imposed in order to reduce the
large parameter space arising from the unknown mechanism of SUSY
breaking. The results of this paper are interpreted within the gravity
mediated SUSY breaking scenario (an expansion of the mSUGRA
model~\cite{Nilles:1984ge, Martin:1997ns, *Arnowitt:1997um}) and the
anomaly mediated SUSY breaking scenario (AMSB~\cite{Randall:1998uk,
  Giudice:1998xp}). While the former scenario offers an elegant way of
explaining the breaking mechanism without additional fields or
interactions, gravity mediated models in general suffer from large
flavour changing neutral currents which must be removed by
fine-tuning. Within the gravity mediated framework, we will consider
both the so-called gaugino-scenario and higgsino-scenario which have
different chargino production mechanisms and cross-sections. The AMSB
scenario exploits the fact that rescaling anomalies in the
supergravity Lagrangian always give rise to soft mass parameters and
thus anomalies always contribute to the SUSY breaking mechanism.

If charginos exist and are sufficiently light, they can be
pair-produced at LEP and may decay mainly via a virtual \W{} boson
into neutralinos and fermions\footnote{The probability for decays via
  sfermions increases with decreasing sfermion masses.}. The kinematic
properties of the final state particles, and therefore the event
topologies, depend on the mass difference between chargino and
neutralino, \DM{}{}. Chargino search strategies can be roughly
categorised by the value of \DM{}{}. For large mass differences
($\DM{}{}\gtrsim5\,\GeV$) the final state fermions are quarks or
energetic leptons and the signal events can be identified by missing
energy (carried away by the neutralinos) plus jets or isolated
leptons. For very small mass differences ($\DM{}{}\lesssim\M{\pi}$)
the chargino lifetime increases and the chargino decay vertex should
be well separated from the interaction point. Heavy stable charged
particles and/or secondary vertices are the signatures for this
scenario. Searches for all these topologies were carried out by all
LEP experiments~\cite{Abbiendi:1999ar, *Ackerstaff:1998si,
  Barate:1999fs, *Barate:1999gm, Abreu:2000nv, *Abreu:2001mc,
  Acciarri:1999km, *Acciarri:1999bw}, but no evidence for a signal
was found.

In this note, scenarios with a mass difference in the intermediate
range ($0.17\,\GeV\lesssim\DM{}{}\lesssim5\,\GeV$) are investigated.
The searches for heavy stable charged particles have high efficiency
and hence still some sensitivity to the \DM{}{} region between $M_\pi$
and $0.17\,\GeV$. In the intermediate \DM{}{} regime the signal final
state is characterised by very little hadronic or leptonic activity
accompanied by a large amount of missing energy. Events from
two-photon processes give rise to a very large background which is
difficult to suppress.  However, due to the large predicted
cross-sections for chargino pair-production, the search can be
restricted to events which also have an energetic photon from initial
state radiation (ISR). These are distinguishable from two-photon
background~\cite{Chen:1996yu, *Chen:1999yf}. This method was proposed
in~\cite{Riles:1990hd} for the search for heavy charged leptons which
are nearly mass degenerate with their neutrino. Recent search results
from other LEP experiments, using this method, are reported
in~\cite{Heister:2002mn, Abreu:2000as, Acciarri:2000wy}.

Section~\ref{sec:detector} of this note describes the OPAL detector
while Section~\ref{sec:mc} details the simulation of both signal and
background events. Section~\ref{sec:analysis} deals with event
selection and Section~\ref{sec:systematics} with systematic
uncertainties. No evidence for chargino pair-production was found and
the search results are interpreted in Section~\ref{sec:results} in
terms of upper limits on the production cross-section and in terms of
lower mass-limits within the MSSM framework.


\section{The OPAL Detector}
\label{sec:detector}

The OPAL detector is described in detail in~\cite{Ahmet:1991eg,
  *Anderson:1998xw, *Anderson:1994ve, Aguillion:1998pz}.  It was a
multipurpose apparatus with almost complete solid angle coverage. The
central detector consisted of two layers of silicon strip detectors
and a system of gas-filled tracking chambers in a 0.435~T solenoidal
magnetic field which was parallel to the beam axis\footnote{In the
  OPAL right-handed coordinate system the $x$-axis points towards the
  centre of the LEP ring, the $y$-axis points upwards and the $z$-axis
  points in direction of the electron beam. The polar angle $\theta$
  and the azimuthal angle $\phi$ are defined with respect to the
  $z$-axis and $x$-axis, respectively.}.  The barrel time-of-flight
scintillator bars were located outside the solenoid, and the end-caps
were equipped with scintillating tiles~\cite{Aguillion:1998pz}. The
scintillator systems were surrounded by a lead-glass electromagnetic
calorimeter (ECAL), which gave hermetic coverage in the region
$|\cos\theta|<0.984$. The magnet return yoke was instrumented for
hadron calorimetry. The forward region was covered by electromagnetic
calorimeters, namely the silicon-tungsten calorimeter (SW,
\mbox{$25\,\mrad<\theta<59\,\mrad$}), the forward detector (FD,
\mbox{$47\,\mrad<\theta\lesssim 160\,\mrad$}) and the gamma catcher
(GC, \mbox{$143\,\mrad<\theta<193\,\mrad$}). Taking into account
material to shield the detector from accelerator related backgrounds,
the detector was sensitive down to $\theta_\MRM{min}=33\,\mrad$.


\section{Data Sample and Monte Carlo Simulation}
\label{sec:mc}

\subsection{The Data Sample}
\label{subsec:data_sample}
The data analysed in this note were taken in 1998, 1999 and 2000 at
centre-of-mass energies $\sqrt{s}$ between 188 and 209 \GeV. The
search is based on a total of 569.9~$\mathrm{pb}^{-1}$ of data for
which all relevant detector components were fully operational. The
luminosity weighted mean centre-of-mass energy and the integrated
luminosity of each energy bin is listed in Table~\ref{tab:lumi}.

\begin{table}[htbp]
  \centering
    \begin{tabular}[H]{|c|c|c|c|}  \hline
year & $\sqrt{s}$ bin range & $\langle\sqrt{s}\rangle$ & $\int\mathcal{L}\ud t$ \\
     &  $[\GeV]$            & $[\GeV]$                 & $[\mathrm{pb}^{-1}]$   \\ \hline
1998 & 188.0 -- 190.5        & 188.6 & 167.6 \\ \hline
1999 & 190.5 -- 194.0        & 191.6 &  27.7 \\
1999 & 194.0 -- 199.0        & 195.5 &  71.2 \\
1999 & 199.0 -- 201.0        & 199.5 &  72.1 \\
1999 & 201.0 -- 202.5        & 201.6 &  33.9 \\ \hline
2000 & 203.5 -- 204.5        & 204.0 &   7.7 \\
2000 & 204.5 -- 205.5        & 205.2 &  64.9 \\
2000 & 205.5 -- 206.5        & 206.3 &  58.6 \\
2000 & 206.5 -- 207.5        & 206.7 &  58.9 \\
2000 &  $>207.5$            & 208.1 &   7.3 \\ \hline
     &                      &       & 569.9 \\ \hline
    \end{tabular}
    \caption{The luminosity weighted mean centre-of-mass energy and
     integrated luminosity of each energy bin.}
    \label{tab:lumi}
\end{table}

\subsection{Signal Simulation}
\label{subsec:signal-simulation}
The SUSYGEN~\cite{Katsanevas:1998fb} generator was used to simulate
the signal events. In this program initial state corrections are
incorporated using a factorised ``radiator formula'' (REMT by
Kleiss~\cite{Beenakker:1996kt}) where exponentiation of higher orders
and the transverse momentum distribution of the photon have been
implemented. The generated chargino masses \MCHI{\pm}{1} ranged from
45 \GeV to 97.5 \GeV. The step size was 5~\GeV for
$\MCHI{\pm}{1}\leq80\,\GeV$. Above this mass the step size was reduced
to $2.5\,\GeV$. The simulated mass differences were \DM{}{}=0.17, 0.2,
0.3, 0.4, 0.5, 1.0, 1.5, 2.0, 3.0, 4.0 and 5.0 \GeV. The points were
chosen to give a (\MCHI{\pm}{1}, \DM{}{}) grid covering all points at
which a reasonable signal efficiency was expected, also taking into
account existing chargino mass limits from LEP 1. A total of 500
signal events were generated for each grid point at $\sqrt{s}=$189,
192, 196, 200, 202, 204, 205, 206, 207 and 208 \GeV, and for both the
higgsino-like and gaugino-like scenarios, in view of their different
ISR spectrum shapes. The signal events were required to be accompanied
by one ISR photon within the ECAL acceptance $|\cos\theta|<0.984$. The
transverse momentum of the ISR photon was required to be larger than
$0.025\,\SQRTS$.

The chargino decay width implemented in SUSYGEN was modified. New
hadronic chargino decay channels
($\tilde{\chi}^-_1\to\pi^-\tilde{\chi}^0_1$,
$\tilde{\chi}^-_1\to\pi^-\pi^0\tilde{\chi}^0_1$ and
$\tilde{\chi}^-_1\to\pi^-\pi^-\pi^+\tilde{\chi}^0_1$) were added
reflecting analytical calculations described in~\cite{Chen:1996yu,
  *Chen:1999yf, Thomas:1998wy}.  The hadronic \CHI{\pm}{1} decays are
modelled using these channels as long as the sum of their partial
widths is larger than the width of the tree level
$\tilde{\chi}^\pm_1\to \qqp \tilde{\chi}^0_1$ decay. The
$\tilde{\chi}^\pm_1\to \qqp \tilde{\chi}^0_1$ decay mode dominates for
$\DM{}{}\gtrsim1.5\,\GeV$.


\subsection{Background Simulation}
The following background processes were considered: lepton pairs
($\eeZ\ell\bar{\ell}$), multi-hadronic ($\eeZ\MRM{q}\bar{\MRM{q}}$),
four-fermion ($\mathrm{e}^+\mathrm{e}^-\to
\mathrm{f}\bar{\mathrm{f}}\mathrm{f}\bar{\mathrm{f}}$) and two-photon
($\mathrm{e}^+\mathrm{e}^-\to\mathrm{e}^+\mathrm{e}^-\gamma\gamma
\to\mathrm{e}^+\mathrm{e}^-X$) processes.

KK2F~\cite{Jadach:1999vf} was used to simulate $\tau^+\tau^-(\gamma)$,
$\mu^+\mu^-(\gamma)$ and multi-hadronic final states.  NUNUGPV
\cite{Montagna:1998ce} was used to generate $\nu\bar{\nu}(\gamma)$
events. TEEGG~\cite{Karlen:1987vk} and BHWIDE~\cite{Jadach:1997nk}
were used to simulate $\MRM{e}^+\MRM{e}^-(\gamma)$ events.
RADCOR~\cite{Berends:1981px} was used to simulate photon pair final
states.  KORALW~\cite{Skrzypek:1996wd, *Skrzypek:1996ur} and
grc4f~\cite{Fujimoto:1997wj} were used to generate four-fermion
events.  The latter was used for all four-fermion final states
involving $\MRM{e}^+\MRM{e}^-$ pairs.  Both KORALW and grc4f simulate
ISR, but only the KORALW samples include ISR photons with transverse
momentum.

PHOJET~\cite{Engel:1995vs, *Engel:1995sb, *Engel:1996yd},
HERWIG~\cite{Marchesini:1992ch} and
VERMASEREN~\cite{Vermaseren:1983cz} were used to simulate two-photon
collisions resulting in electron pair or hadronic final states.  None
of these Monte Carlo generators include QED radiative corrections. The
generator of Berends, Darverveldt and Kleiss
(BDK)~\cite{Berends:1985ge, *Berends:1986if, *Berends:1986ig,
  *Berends:1986ih} is able to handle these corrections. The program
was written to simulate $\mathrm{e}^+\mathrm{e}^-
\to\mathrm{e}^+\mathrm{e}^-\gamma^\star\gamma^\star(\gamma)
\to\mathrm{e}^+\mathrm{e}^-\mu^+\mu^-(\gamma)$ events. For this
analysis BDK was modified to allow the generation of $\tau$ final
states. While for the $\tau$ samples no cuts were applied at generator
level, in the case of the $\mu$ samples an event was accepted if it
satisfied one of the following three requirements: it contained at
least one particle with $|\cos\theta|<0.975$ and a transverse momentum
(with respect to the beam axis) larger than 4.0~\GeV or it contained
two particles with $|\cos\theta|<0.975$ and the total missing
transverse momentum (with respect to the beam axis) of the event was
larger than 1.0~\GeV or it contained more than two particles with
$|\cos\theta|<0.975$. Since BDK includes only the QED coupling of
photons, it cannot be used to describe all features of two-photon
processes, e.g.  the hadron-like structure of photons.  Therefore no
attempt was made to fully model the background from radiative hadronic
two-photon events.  However, BDK was used to verify that the veto on
these events operates efficiently. The veto cuts (see
Section~\ref{subsubsec:veto}) depend predominantly on the
characteristics of the ISR photon and the beam electrons\footnote{Both
  electrons and positrons are referred to as electrons.} and not on
the properties of the two-photon system.

Unless specified, JETSET 7.4~\cite{Sjostrand:1986ys,
  *Sjostrand:1994yw} was used for the fragmentation of final states
involving quarks. The number of events generated was typically well
above the number expected in data. All Monte Carlo events generated
were passed through a complete simulation of the OPAL
detector~\cite{Allison:1992bf}, and processed in the same way as the
data.


\section{The Analysis}
\label{sec:analysis}
Signal events from the process
$\ee\to\CHI{+}{1}\CHI{-}{1}\gamma\to\CHI{0}{1}\CHI{0}{1}\gamma + X$
are characterised by an energetic photon accompanied by large missing
energy and momentum and little hadronic or leptonic activity.  The
photon increases the total visible energy of the event and guarantees
a high trigger efficiency ($\sim100\%$). Since only a small fraction
of signal events are accompanied by a hard ISR photon the signal
efficiency is expected to be rather low. However, given the high
integrated luminosities of the LEP~2 data samples a signal would still
be observable.

Radiative two-photon events $\ee\to\ee\gamma+ X$ can lead to a similar
topology to the signal. However, in this case a final state electron
can also be observed if it has a polar angle larger than
$\theta_{\MRM{min}}$. This angle can be related to the minimum
transverse energy $E_\MRM{T}^{\gamma\MRM{min}}$ of the detected
photon~\cite{Gunion:2001fu}:
\begin{eqnarray}
\label{eq:veto}
E_\MRM{T}^\gamma\gtrsim{\ETg}^{\MRM{min}}=\SQRTS\frac{\sin\theta_{\MRM{min}}}
{1+\sin\theta_{\MRM{min}}}.
\end{eqnarray}
Therefore, requiring a photon with $\ETg > {\ETg}^{\MRM{min}}$ in the
final state allows an efficient reduction of two-photon processes by
vetoing events with a scattered beam electron. The expression in
Eq.~(\ref{eq:veto}) is an approximation, which allows for resolution
effects. 

Unless specified, the same quality requirements for tracks and ECAL
clusters were used as in~\cite{Abbiendi:1999sx}.  Double-counting of
energy between tracks and calorimeter clusters was corrected as
in~\cite{Ackerstaff:1998ng}. This procedure results in a set of
reconstructed particles. Photons were identified by an algorithm
similar to that described in detail in~\cite{Ackerstaff:1998id,
  *Abbiendi:1999aa}. Photon candidates were identified as one of three
types: 
\begin{itemize}
\item A photon candidate was defined as an ECAL cluster including at
  least two lead-glass blocks, and with no reconstructed track
  compatible with being associated with the cluster. Furthermore, the
  energy of additional tracks and clusters in a $15^\circ$ half-angle
  cone defined by the photon direction had to be less than 2~\GeV. The
  photon candidate was accepted if $|\cos\theta^\gamma|<0.976$. If
  several photon candidates were found they were ordered according to
  their energy beginning with the highest energy.
\item Two-track photon conversions were selected using an artificial
  neural network.
\item Conversions where only a single track was reconstructed were
  defined as an ECAL cluster associated with a reconstructed track
  which was consistent with pointing to the primary vertex. 
\end{itemize}
For this analysis only non-converted ISR photon candidates were
considered.


\subsection{Event Selection}

\subsubsection{Pre-Selection}
Events were selected if a photon candidate was found in the ECAL. The
photon transverse energy had to be larger than 5 \GeV. In order to
reduce the background the reduced visible energy, \Evis, defined as
the difference between total visible energy and the photon candidate
energy, $E^\gamma$, was required to be less than $0.35\,\SQRTS$.
High-multiplicity events were rejected by requiring that the number of
tracks passing the track quality cuts~\cite{Abbiendi:1999sx},
excluding tracks which had been assigned to a photon conversion, had
to be at least two and at most ten.  In addition beam-gas and
beam-wall interactions and cosmic background were rejected by a veto
described in~\cite{Abbiendi:2000hh} using ECAL cluster shapes,
information from the time-of-flight system, the tile end-cap
scintillators, the muon chambers and the hadronic calorimeter. After
this rather loose pre-selection all events had to pass the veto on
two-photon events (V), a more refined cleaning selection (C), and a
final selection (F). If events contained more than one photon
candidate the chain of cuts was repeated for each photon separately
starting with the one with the highest energy until one candidate was
found passing all cuts or all candidates failed one cut.

\subsubsection{The Two-Photon Veto}
\label{subsubsec:veto}
Two-photon events were vetoed by additional energy deposited in the
detector from a final state electron (``two-photon veto'').

\begin{itemize}
\item[V1] The photon candidate had to pass the \ETg requirement Eq.
      (\ref{eq:veto}). With a minimal accessible electron polar angle of
      $\theta_\mathrm{min}=33$ mrad the transverse
      photon energy was required to satisfy ${\ETg} \geqslant 0.0319\SQRTS$.
\item[V2] The sum of energies, $E_\mathrm{fwd}$, detected in the forward
      detectors SW, FD, GC on either side had to be smaller than 5 \GeV.
\end{itemize}
From the two-photon samples produced with the BDK generator a
two-photon veto efficiency of $\varepsilon^{\gamma\gamma}>99.9\%$ was
derived.

\subsubsection{Cleaning Cuts}
\label{subsubsec:cleaning-cuts}
\begin{figure}
    \begin{center}
    \includegraphics[width=.95\textwidth]{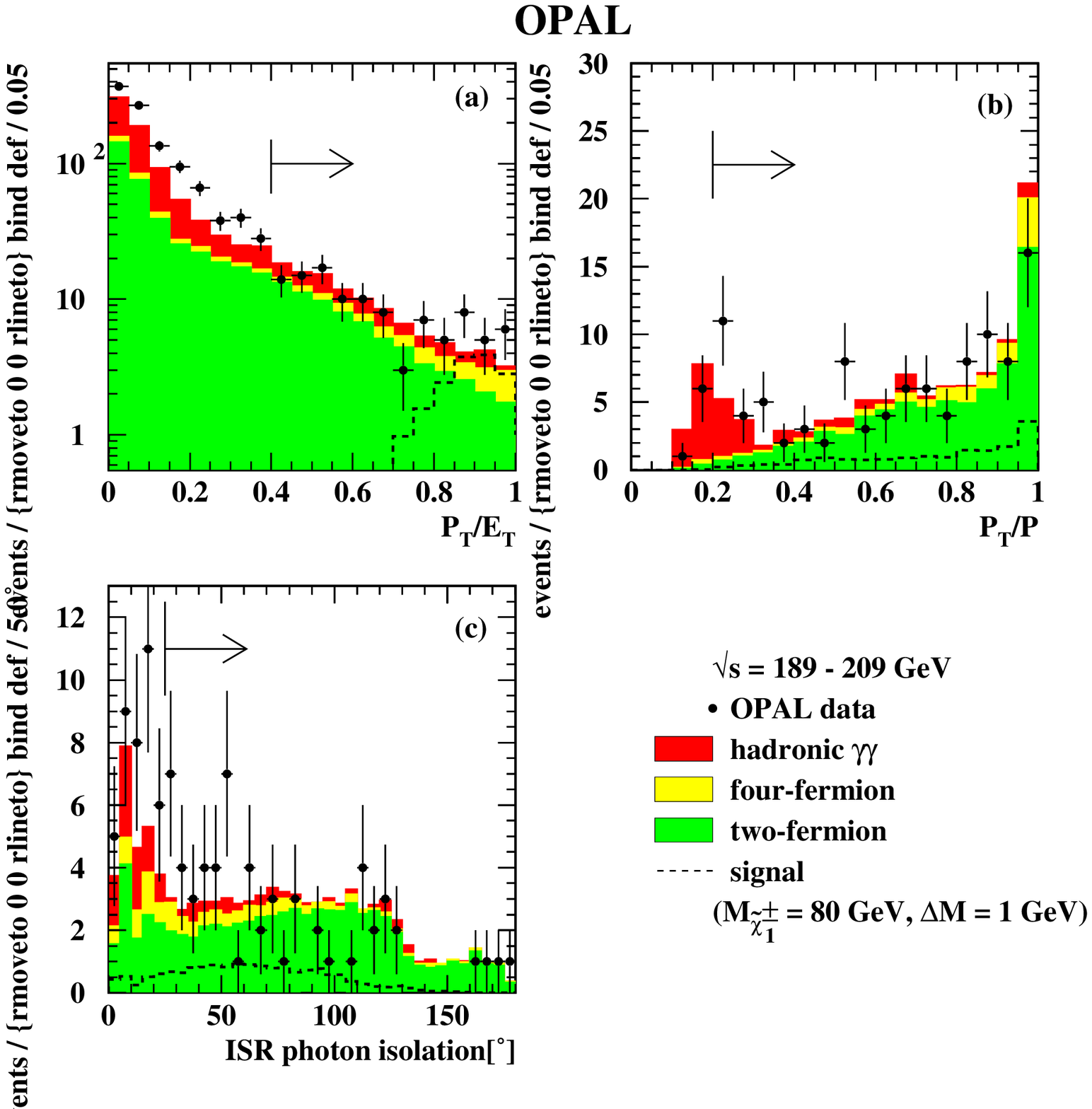}
    \caption{Variables used as cleaning cuts. The OPAL data are
      indicated by points with error bars (statistical error) and the
      background distributions by the histograms.  Signal
      distributions are plotted as dashed lines and correspond to a
      chargino with $\M{\CHI{\pm}{1}}= 80\,\GeV$ and
      $\DM{}{}=1\,\GeV$. The accumulated events for $\sqrt{s}=$189 --
      209 \GeV are shown.  Each plot shows the distribution after the
      cuts on the preceeding variables. The arrows indicate the
      accepted regions.}
    \label{fig:cleaning-cuts}
    \end{center}
\end{figure}

After the pre-selection and the two-photon veto, the remaining events
in the background sample are mainly hadronic two-photon events with a
faked or misinterpreted ISR photon, or from two-fermion events.
Figure~\ref{fig:cleaning-cuts} shows the distributions of the
variables used for the cleaning cuts against this background. The cut
variables are based mainly on the properties of the reconstructed
particles: The total transverse momentum $P_\mathrm{T}$ is the
component of the total momentum $P$ (including the ISR candidate)
transverse to the beam axis. The transverse visible energy
$E_\mathrm{T}$ was calculated by summing up the absolute transverse
momenta (with respect to the beam axis) of all reconstructed
particles.

\begin{itemize}
\item[C1] Most of the events from hadronic two-photon processes and
  two-fermion events were removed by demanding that the ratio of the
  total transverse momentum to the transverse energy
  $P_\mathrm{T}/E_\mathrm{T}$ is larger than 0.4.
\item[C2] A further reduction of the hadronic two-photon background
  was achieved by requiring that the ratio of the total transverse
  momentum to the total momentum $P_\mathrm{T}/P$ is larger than 0.2.
\item[C3] The definition of a photon candidate was further restricted
  with the following cuts.  The relative photon energy measurement
  error had to be less than $30\%$.  In addition the angle between the
  photon candidate and any track or ECAL cluster was required to be
  larger then $25^\circ$.
\end{itemize}

As shown in Figure~\ref{fig:cleaning-cuts} the agreement between data
and Monte Carlo is poor, but this is expected due to the unmodelled
ISR spectrum in most of the Monte Carlo samples and the uncertainty on
the two-photon cross-sections. However, the largest disagreement lies
outside the accepted regions.

\subsubsection{The Final Selection}
\label{subsubsec:final}
\begin{figure}
    \begin{center}
    \includegraphics[width=.95\textwidth]{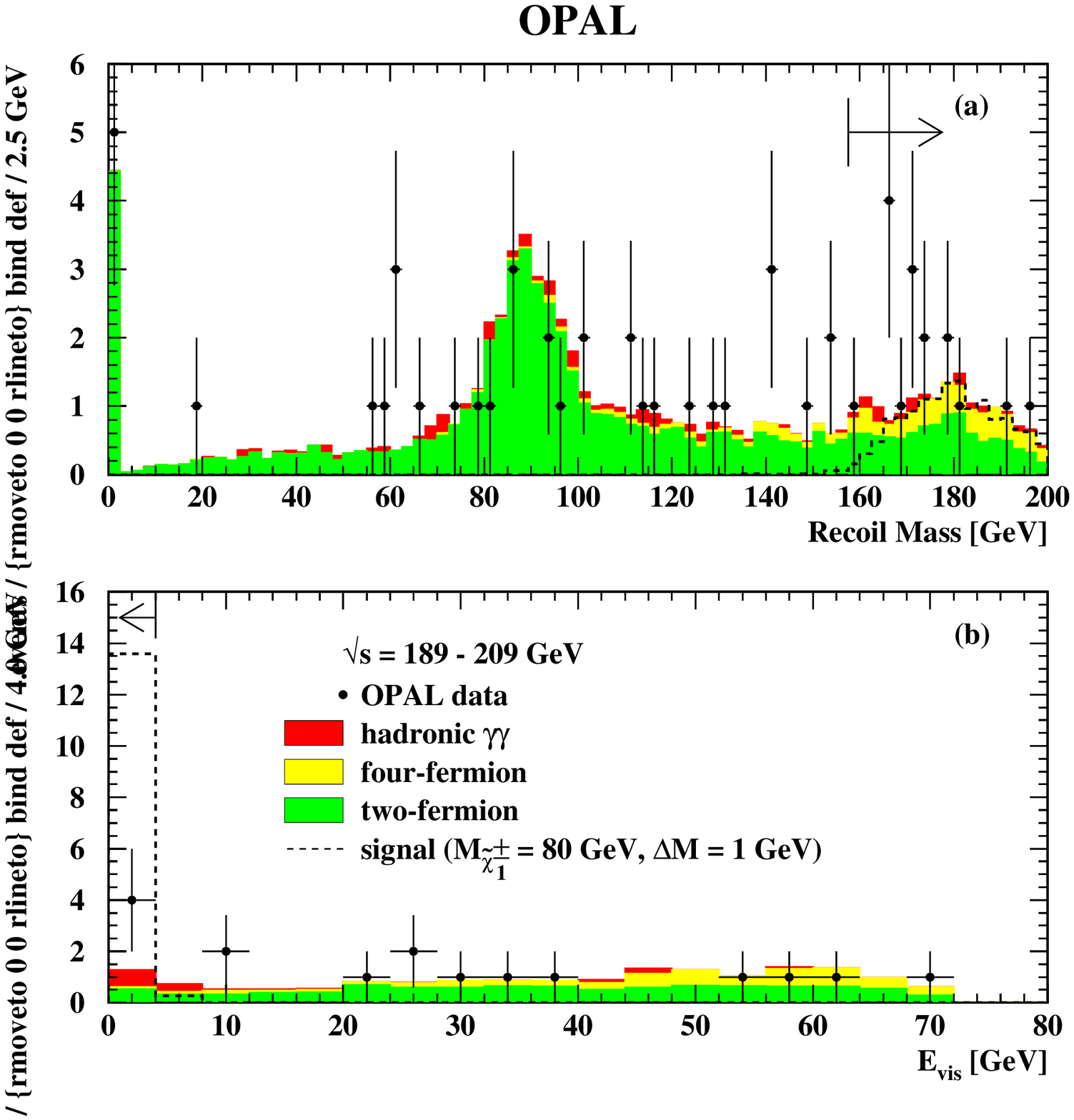}
    \caption{Variables used in the final event selection. The OPAL
      data are indicated by points with error bars (statistical error)
      and the background distributions by the histograms.  Signal
      distributions are plotted as dashed lines and correspond to a
      chargino with $\M{\CHI{\pm}{1}}= 80\,\GeV$ and
      $\DM{}{}=1\,\GeV$. The accumulated events for $\sqrt{s}=$189 --
      209 \GeV are shown.  Each plot shows the distribution after the
      cuts on the preceeding variables. The arrows indicate the
      accepted regions.}
    \label{fig:final-cuts}
    \end{center}
\end{figure}

To optimise the signal to background ratio, the final two cuts are a 
function of the point in the (\M{\CHI{\pm}{1}}, \DM{}{}) space which 
is being tested.
\begin{itemize}
\item[F1] The ISR photon candidate recoil mass is defined as
     $M_\mathrm{rec}=\SQRTS(1-2E^\gamma/\SQRTS)^\frac{1}{2}$.
     The recoil mass should be at least twice the tested chargino
     mass
     \MCHI{\pm}{1}. Taking resolution effects into account the cut on the 
     recoil mass was set to $2\times\MCHI{\pm}{1}-2.5\,\GeV$.
   \item[F2] The visible energy \Evis depends to first order on the
     tested mass difference \DM{}{}. In order to retain the maximum
     signal efficiency, \Evis was required to be smaller than
     $4\times\DM{}{}$.
\end{itemize}
The recoil mass distribution after cut C3 and the visible energy
distribution after cut F1 are shown in Figure~\ref{fig:final-cuts}.

\subsection{Selection Results}
\label{subsec:selection-results}

The numbers of observed events and expected background events after
each cut for the full available data set ($\sqrt{s}=$189 -- 209 \GeV)
are summarised in Table~\ref{tab:cut-flow}. The disagreement between
data and Monte Carlo during the pre-selection vanishes after the
two-photon veto. After the final cut F2 no excess over the SM
background is observed.

\begin{table}[htbp]
  \centering
  \begin{tabular}[H]{|c|ccccc|}   \hline
     &  two-fermion & four-fermion & two-photon & total bkg. & data \\ \hline
   P &  $584.7\pm8.4$  & $69.2\pm1.1$   & $923.0\pm13.5$ &$1580\pm16$    & 2432\\ \hline  
  V1 &  $561.0\pm8.2$  & $63.1\pm1.1$   & $607.9\pm9.9$  &$1235\pm13$    & 1881\\    
  V2 &  $427.8\pm7.3$  & $47.4\pm0.9$   & $393.2\pm8.0$  &$870.7\pm10.9$ & 1150\\ \hline   
  C1 &  $71.0\pm0.9$ & $12.9\pm0.4$ & $24.4\pm1.3$ &$108.3\pm1.6$  & 108\\
  C2 &  $70.6\pm0.9$ & $12.4\pm0.4$ & $14.4\pm1.0$ &$97.4\pm1.4$ & 101\\
  C3 &  $58.4\pm0.8$ & $8.1\pm0.3$  & $5.4\pm0.6$  &$72.0\pm1.0$ &  52\\ \hline

  F1 &  $9.7\pm0.3$  & $5.5\pm0.3$  & $1.5\pm0.3$  &$16.7\pm0.5$ & 16 \\
  F2 &  $0.5\pm0.1$  & $0.09\pm0.02$& $0.7\pm0.2$  &$1.3\pm0.2$  & 4 \\ \hline
  \end{tabular}
  \caption{Number of expected background events and number of candidates after 
           each cut. The first three rows list the contributions to the expected 
           background from fermion-pair, four-fermion and two-photon events. 
           The hadronic two-photon events passing the two-photon veto are those produced
           with generators without QED radiative corrections. 
           The errors are statistical. All numbers
           correspond to the full analysed data set (569.9
           pb$^{-1}$). For the final cuts F1 and F2 a chargino mass of
           $\M{\CHI{\pm}{1}}=80\,\GeV$ and a mass difference of
           $\DM{}{}=1\,\GeV$ were assumed.} 
  \label{tab:cut-flow}
\end{table}

\begin{figure}
    \begin{center}
    \includegraphics[width=.95\textwidth]{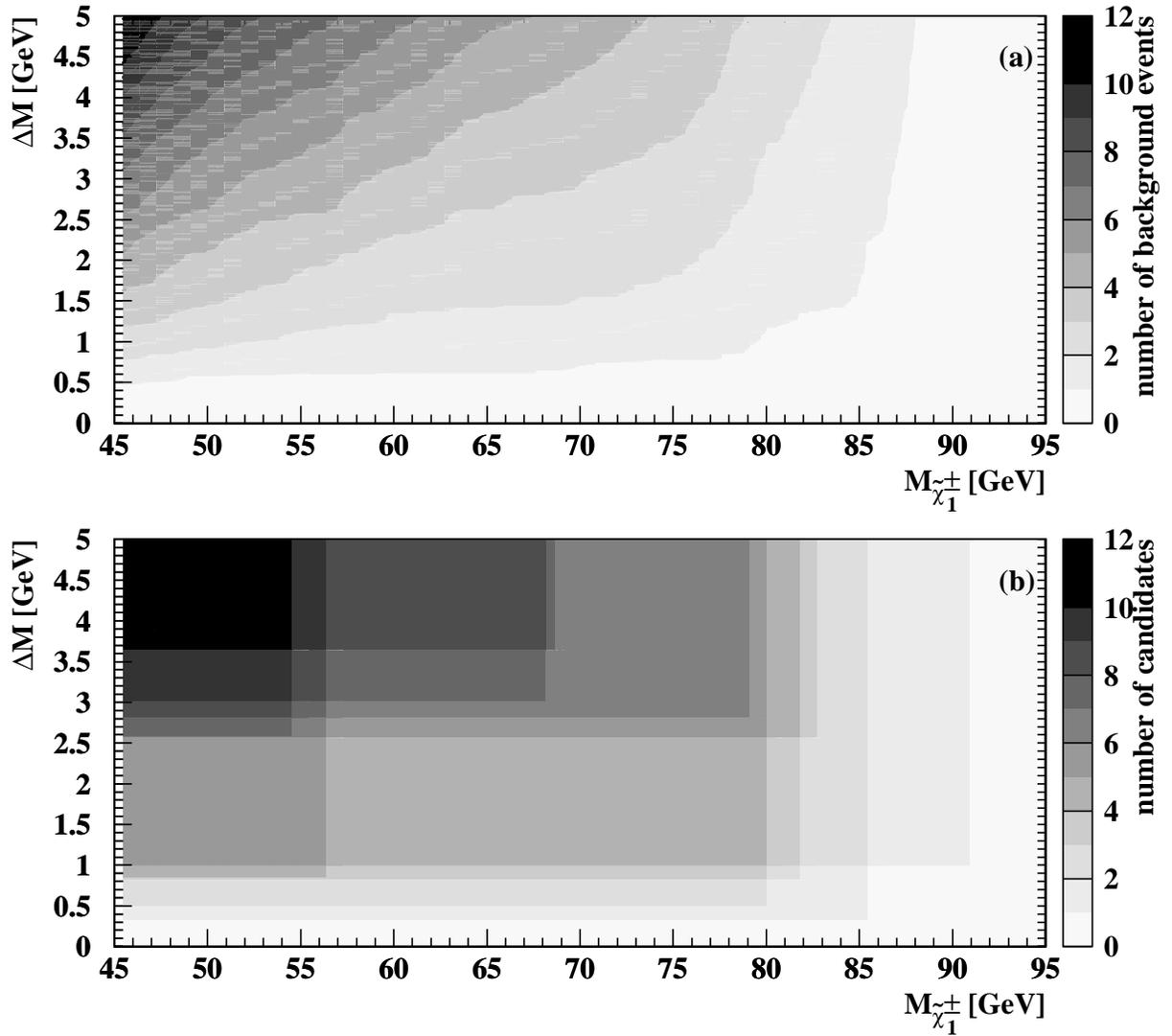}
    \caption{Number of expected background events (a) and number of candidates (b)
      passing the final selection cuts. The accumulated events for
      $\sqrt{s}=$189 -- 209 \GeV are shown.}
    \label{fig:bkgvsdata}
    \end{center}
\end{figure}

The numbers of candidates and expected events after the final cut F2
for a selection of (\M{\CHI{\pm}{1}}, \DM{}{}) grid points are listed
in Table~\ref{tab:grid}. The distribution of candidates and expected
events in the (\M{\CHI{\pm}{1}},\DM{}{}) plane is illustrated in
Figure~\ref{fig:bkgvsdata}.

\begin{table}[htbp]
    \begin{center}
        \begin{tabular}[h]{|cc|cc|} \hline
        $\M{\CHI{\pm}{1}}\,[\GeV]$ & $\DM{}{}\,[\GeV]$ & total bkg. & data \\ \hline
        50    & 4  & $7.3\pm0.5$ & 10 \\
        60    & 3  & $4.1\pm0.4$ &  6 \\
        70    & 2  & $2.4\pm0.3$ &  4 \\
        80    & 1  & $1.3\pm0.2$ &  4 \\
        90    & 0.5 & $0.2\pm0.1$ & 1 \\ \hline 
        \end{tabular}
        \caption{Number of expected background events and number of candidates after 
          the final cut F2 for a selection of (\M{\CHI{\pm}{1}}, \DM{}{})
          points. The errors are statistical. All numbers
          correspond to the full analysed data set (569.9 pb$^{-1}$).}
        \label{tab:grid}
    \end{center}
\end{table}

\begin{figure}
    \begin{center}
    \includegraphics[width=.95\textwidth]{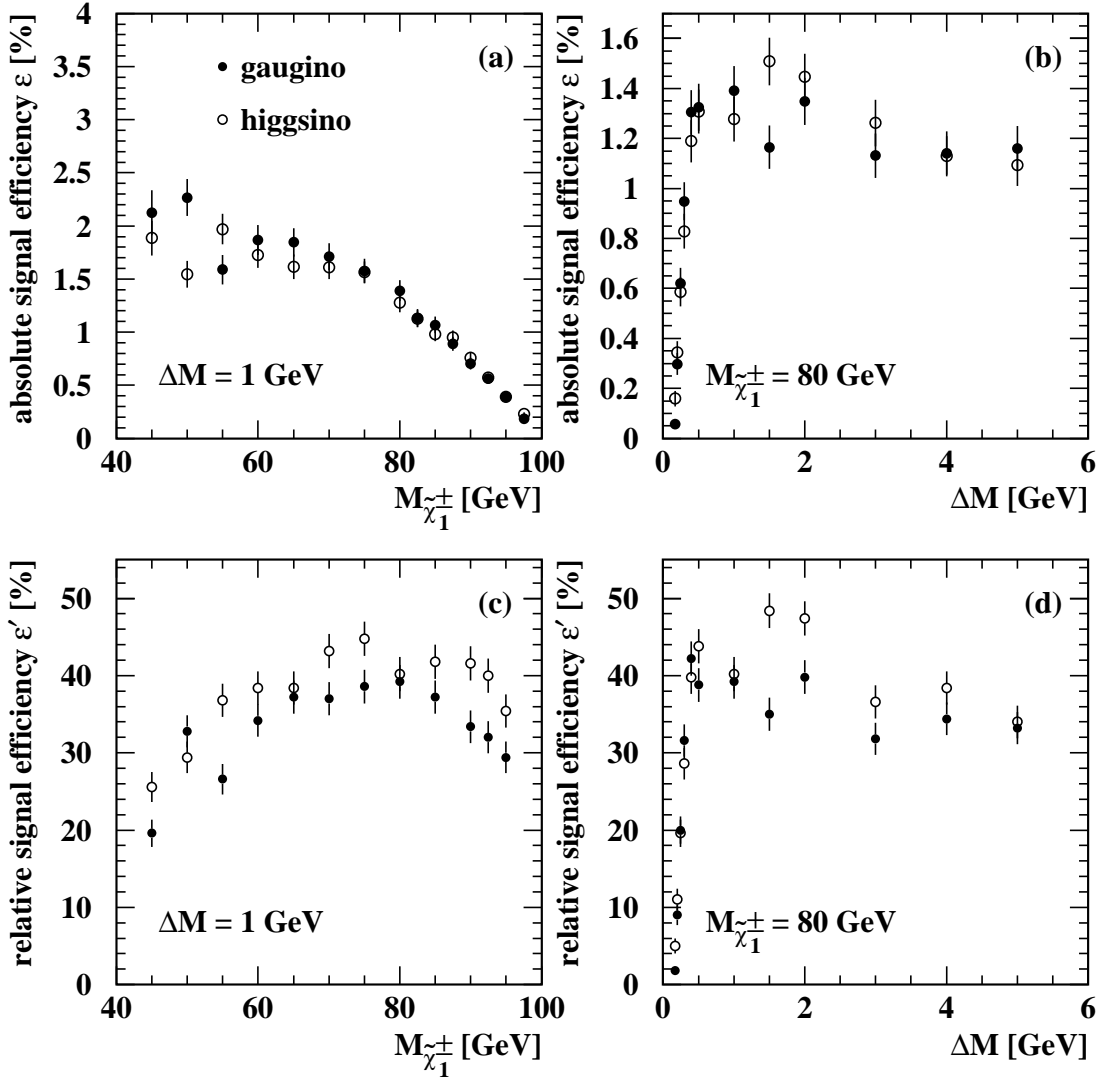}
    \caption{Absolute signal efficiency (number of selected signal
      events normalised to the number of generated events),
      $\varepsilon$, and relative efficiency (number of selected
      signal events normalised to a subset of generated events with
      $\ETg>0.025\sqrt{s}$ and $|\cos\theta^\gamma|<0.985$),
      $\varepsilon^\prime$, at $\sqrt{s}=208\,\GeV$ for a constant
      mass difference $\DM{}{}=1\,\GeV$ ((a)+(c)) and for a constant
      chargino mass $\M{\CHI{\pm}{1}} = 80\,\GeV$ ((b)+(d)).  The
      efficiencies for the gaugino-like scenario are indicated by the
      points.  The circles correspond to the efficiencies in the case
      of a higgsino-like chargino.}
    \label{fig:efficiencies}
    \end{center}
\end{figure}

As an example Figures~\ref{fig:efficiencies} (a) and (b) show the
signal efficiencies $\varepsilon$ at $\sqrt{s}=208\,\GeV$ for a
constant mass difference \DM{}{} and a constant chargino mass
\M{\CHI{\pm}{1}} respectively. The rather low signal efficiencies are
due to the requirement of having a high energy ISR photon within the
detector fiducial region. To obtain the efficiencies for arbitrary
masses and mass differences the measured efficiencies were
interpolated by a spline fit. In order to demonstrate the sensitivity
of the analysis, the ``relative efficiency'', $\varepsilon^\prime$, i.e.
the efficiency for a subset of generated events with
$\ETg>0.025\sqrt{s}$ and $|\cos\theta^\gamma|<0.985$, is presented in
Figure~\ref{fig:efficiencies} (c) and (d).  The significant drop in
efficiency for very small mass differences ($\DM{}{}<0.5\,\GeV$, see
Figures~\ref{fig:efficiencies} (b) and (d)) is due to reduced phase
space.  In this case all of the chargino decay products are particles
with low momentum and most of these events do not pass the track
quality cuts during the pre-selection. Chargino lifetime effects,
which play a significant role in most SUSY scenarios in this very
small \DM{}{} region ($\DM{}{}<0.5\,\GeV$), were not taken into
account. This region is included in the experimental results but is
not used for the theoretical interpretation due to the theoretical
uncertainties.

\section{Systematic Uncertainties}
\label{sec:systematics}
The systematic errors on the signal efficiency arise from several
sources. They are described in the following. The size of the
uncertainties strongly depends on the tested (\M{\CHI{\pm}{1}},
\DM{}{}) point.
\begin{itemize}
\item To take into account the uncertainty due to the limited Monte
  Carlo statistics and the uncertainty introduced when interpolating
  the signal efficiencies the signal efficiencies were randomly
  smeared within their statistical errors and the resulting
  distribution was interpolated. This process was repeated 100 times
  and the absolute values of the differences between the original and
  the smeared fits were averaged. The mean difference in each tested
  (\MCHI{\pm}{1}, \DM{}{}) point was taken as a measure of the
  systematic error.
\item The size of the uncertainty arising from the ISR photon
  simulation was estimated by re-weighting the transverse momentum
  spectrum of ISR photons in SUSYGEN. Event weights were extracted
  from a comparison of the ISR transverse momentum spectra of
  \W{+}\W{-} events produced with the KORALW generator with and
  without QED corrections up to order $\alpha^3$. The maximum
  difference between the interpolated signal efficiencies calculated
  with un-weighted events and the interpolated signal efficiencies
  calculated with re-weighted events was used as the systematic
  uncertainty for all grid points and centre-of-mass energies.
\item Another systematic uncertainty arises from the modelling of the
  ECAL energy scale and the ECAL energy resolution which directly
  affect the reconstructed ISR photon candidate. The size of the
  data-Monte Carlo discrepancy for the energy scale and the energy
  resolution was studied using $\ee\to\ee$ events from high energy
  data as well as \Z{0} calibration data.\\ 
  To account for energy scale uncertainty, the ISR photon energy was
  shifted by $\pm1\%$. The resulting interpolated efficiencies were
  compared with the original efficiency spline. The maximum deviation
  in each tested (\MCHI{\pm}{1}, \DM{}{}) point from the
  original spline was used as a measure of the systematic error.\\
  To study the ECAL resolution effects the difference between true and
  measured photon energy divided by the measured energy error was
  calculated. This pull distribution was broadened by $\pm20\%$.
  Event weights were extracted from the ratios of the original and
  broadened pull distributions. The resulting signal efficiencies were
  interpolated and the fits were compared with the original efficiency
  spline.  Again, the maximum deviation in each tested (\MCHI{\pm}{1},
  \DM{}{}) point was taken as the systematic uncertainty.
\item In addition, the systematic uncertainty on modelling the cut
  variables were studied.  The cut on the visible energy, F2, was of
  particular importance. It was shifted by $\pm5\%$ and the resulting
  efficiencies were compared with the original ones. The maximum
  deviation in each tested (\MCHI{\pm}{1}, \DM{}{}) point from the
  original spline was assumed as systematic error. 
\item The effect of changing the decay model as described in
  Section~\ref{subsec:signal-simulation} was studied. It was found to
  be small and therefore was neglected in the interpretation.
\end{itemize}

The following sources of systematic uncertainties on the number of
expected background events were investigated. The limited Monte Carlo
statistics was taken into account.  The systematic uncertainties due
to the Monte Carlo modelling of the ECAL energy and the ECAL energy
resolution as well as the modelling of the cut variables were
estimated as described above. As an example
Table~\ref{tab:systematics} gives an overview on the size of
systematic uncertainties for the signal efficiency and the number of
expected background events at $\M{\CHI{\pm}{1}}= 80\,\GeV$ and
$\DM{}{}=1\,\GeV$.

\begin{table}[htbp]
  \centering
  \begin{tabular}[h]{|c|c|c|} \hline
                   & signal efficiency ($\sqrt{s}=208\,\GeV$)  &
                   background ($\sqrt{s}=189-208\,\GeV$)  \\ \hline
central value      &    $1.3\%$         &  1.3        \\ \hline\hline
statistical error/ &                    &             \\
interpolation      &    \RB{$\pm1.5\%$}    &  \RB{$\pm0.2$}   \\ \hline 
ISR modelling      &    $\pm22.3\%$        &  --         \\ \hline
ECAL resolution    &    $\pm7.6\%$         &  $\pm0.6$        \\ \hline 
energy scale       &    $\pm0.8\%$         &  $\pm0.1$        \\ \hline
cut modelling      &    $\pm1.5\%$         &  $\pm0.1$           \\  \hline\hline
total              &    $\pm23.7\%$        &  $\pm0.6$      \\ \hline 
  \end{tabular}
  \caption{The left column lists the signal efficiency and relative
                   statistical and systematic 
                   uncertainties at $\sqrt{s}=208\,\GeV$ for
    $\M{\CHI{\pm}{1}}=80\,\GeV$ and $\DM{}{}=1\,\GeV$ after the final
    cut F2. The right column shows the total number of
    expected background events and the absolute statistical and
    systematic uncertainties for $\M{\CHI{\pm}{1}}=80\,\GeV$ and
    $\DM{}{}=1\,\GeV$ after the final cut F2.}
  \label{tab:systematics}
\end{table}

For the two-photon veto a cut on the amount of energy deposited in the
forward detectors was applied (see Section~\ref{subsubsec:veto}).
This energy could also have been deposited by accelerator related
activity which was not included in the Monte Carlo simulation. The
assumed integrated luminosity used for the limit calculation was
reduced by $\sim3\%$, estimated from a sample of random beam crossing
events, to account for this.

\section{Results}
\label{sec:results}
No evidence was observed for \CHI{+}{1}\CHI{-}{1} production. Exclusion
regions and limits were determined for various scenarios using the 
likelihood ratio method described in~\cite{Junk:1999kv}. The method is
able to combine results from different search channels. This feature 
was used for the combination of the results at different centre-of-mass 
energies. Systematic uncertainties were taken into account using a Monte 
Carlo technique allowing for a correct treatment of correlated errors.

\subsection{Cross-section Limits}
\label{subsec:xs-limits}
\begin{figure}
    \begin{center}
    \includegraphics[width=.95\textwidth]{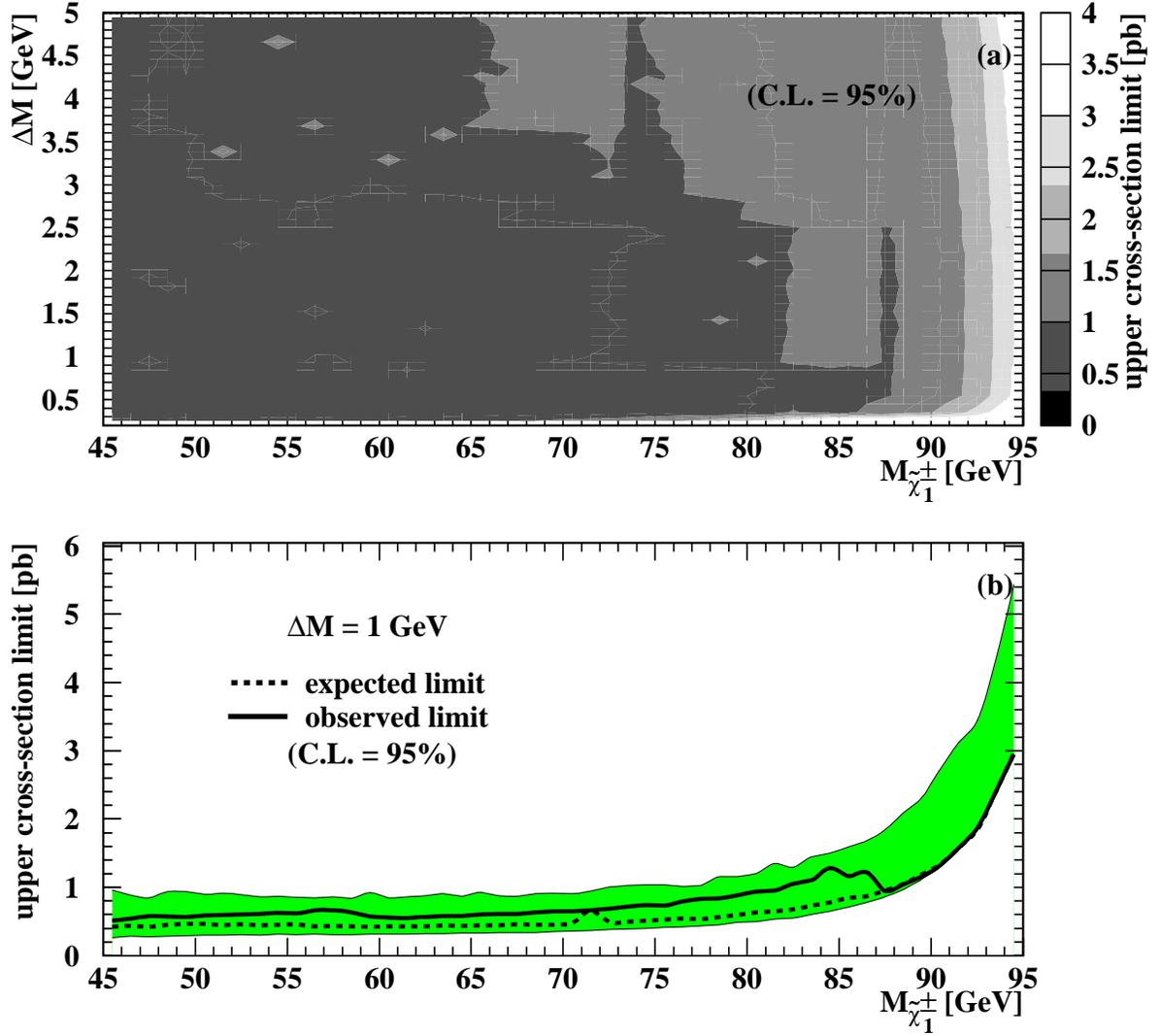}
    \caption{Upper signal cross-section limit rescaled to $\sqrt{s}=208\,\GeV$ at the
             95\% confidence level as a function of the mass difference and the
             chargino mass (a) and for a constant mass difference of 
             $\DM{}{}=1\,\GeV$ (b). The band in (b) indicates the
             $\pm2\sigma$ interval around the median expected value.}
    \label{fig:xs-limits}
    \end{center}
\end{figure}

An upper limit for the \CHI{+}{1}\CHI{-}{1} production cross-section
$\sigma$ was derived from the number of expected background events,
the number of observed candidates and the higgsino-like scenario
signal efficiencies. The signal efficiencies of the higgsino-like
scenario are slightly smaller than those of the gaugino-like scenario
and therefore result in more conservative limits.
Figure~\ref{fig:xs-limits}~(a) shows the observed upper cross-section
limits in the (\MCHI{\pm}{1}, \DM{}{}) plane. Production
cross-sections above $\sim$ 0.4 -- 3.7 pb were excluded for chargino
masses up to $\sim 95$ \GeV and for $0.5\,\GeV \leq \DM{}{} \leq
5.0\,\GeV $ at the $95\%$ confidence level (\CL) rescaled to
$\sqrt{s}=208\,\GeV$ assuming the signal cross-section evolution with
$\sqrt{s}$ calculated by SUSYGEN. The expected and observed upper
cross-section limits for \DM{}{}=1 \GeV as a function of the chargino
mass are illustrated in Figure~\ref{fig:xs-limits}~(b).

\subsection{Interpretation within the MSSM}
\label{subsec:MSSM}

Within the supergravity version of the MSSM~\cite{Nilles:1984ge,
  Martin:1997ns, *Arnowitt:1997um}, the chargino and neutralino masses
depend on four parameters: the ratio of the two Higgs vacuum
expectation values, $\tan\beta$; the Higgs mixing parameter, $\mu$ and
the $U(1)$ and $SU(2)$ gaugino mass parameters, $M_1$ and $M_2$
respectively:
\begin{eqnarray}
\MCHI{\pm}{i=1,2}    & = & \MCHI{\pm}{i=1,2}(M_2,\mu,\tan\beta), \nonumber \\
\MCHI{0}{i=1\dots 4} & = & \MCHI{0}{i=1\dots 4}(M_1,M_2,\mu,\tan\beta). \nonumber
\end{eqnarray}
Unless specified, in the following the value of $\tan\beta$ was set to
1.5.  The MSSM can have small mass differences \DM{}{} in the
following two scenarios:
\begin{itemize}
\item If $|\mu| \ll M_2$, the lightest chargino \CHI{\pm}{1} is almost
  a higgsino. In this scenario the lightest chargino and neutralino
  are mass-degenerate (\MCHI{\pm}{1}$\sim$\MCHI{0}{1}$\simeq|\mu|$)
  for very large values of $M_2$. We call this the higgsino-like
  scenario.
\item If $M_2 \ll |\mu|$, the lightest chargino \CHI{\pm}{1} is almost
  a gaugino $\tilde{\mathrm{W}}^\pm$ with $\MCHI{\pm}{1}\sim M_2$.
  The mass of the lightest neutralino \CHI{0}{1} is given by
  $\MCHI{0}{1}\sim\min(M_1,\,M_2)$.  Therefore small mass splittings
  only occur if $M_2<M_1$. If the masses of all gauginos are assumed
  to be identical at the grand unification scale (GUT), then the
  relation between $M_1$ and $M_2$ at the $\mathrm{Z}$-scale is given by:
      \begin{eqnarray}
        \label{eq:M1_M2-ratio1}
        M_1 & = & \tan^2\theta_W M_2 \simeq \dfrac{1}{2} M_2.
      \end{eqnarray}
      In this case the model does not predict a small \DM{}{}
      gaugino-like scenario.  In more general scenarios, Eq.
      (\ref{eq:M1_M2-ratio1}) does not hold and can be generalised
      introducing an arbitrary factor $R_\mathrm{S}$:
      \begin{eqnarray}
        \label{eq:M1_M2-ratio2}
        M_1 & = & R_\mathrm{S} M_2.
      \end{eqnarray}
      Some string theory motivated SUSY scenarios~\cite{Chen:1997ap}
      explicitly predict $R_\mathrm{S} \not=\tan^2\theta_W$. For
      $R_\mathrm{S} \gtrsim 1$ the lightest chargino and neutralino
      are degenerate in mass with $\MCHI{\pm}{1}\sim\MCHI{0}{1}\sim
      M_2$.
\end{itemize}

\begin{figure}
    \begin{center}
    \includegraphics[width=.95\textwidth]{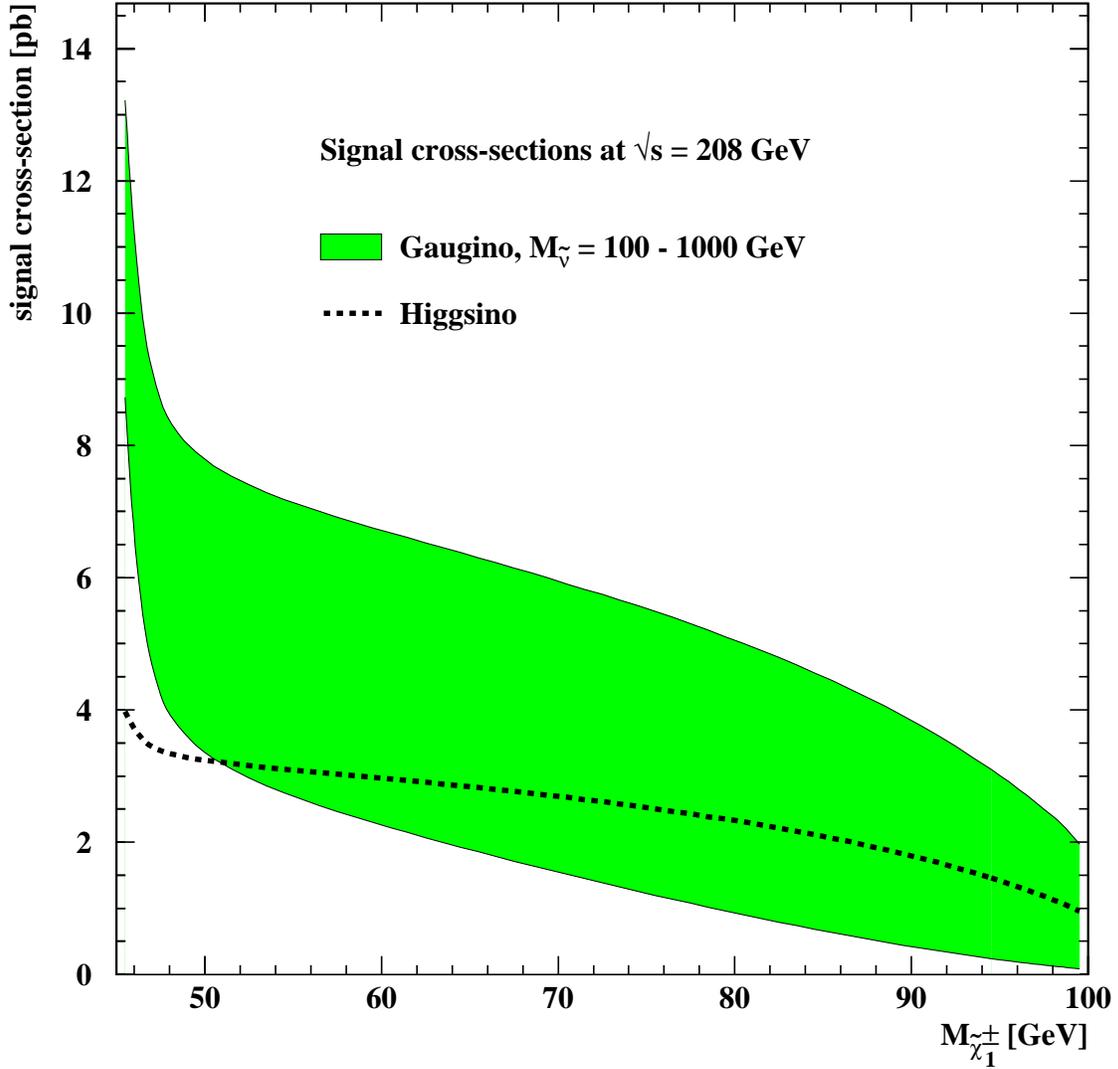}
    \caption{Signal cross-sections as a function of the chargino mass at 
      $\sqrt{s}=208\,\GeV$. The shaded band shows the cross-sections
      in case of a gaugino-like chargino for sneutrino masses ranging
      from 100 to 1000 \GeV. The dashed line corresponds to the
      higgsino-like scenario.}
    \label{fig:xsections}
    \end{center}
\end{figure}

\begin{figure}
    \begin{center}
    \includegraphics[width=.95\textwidth]{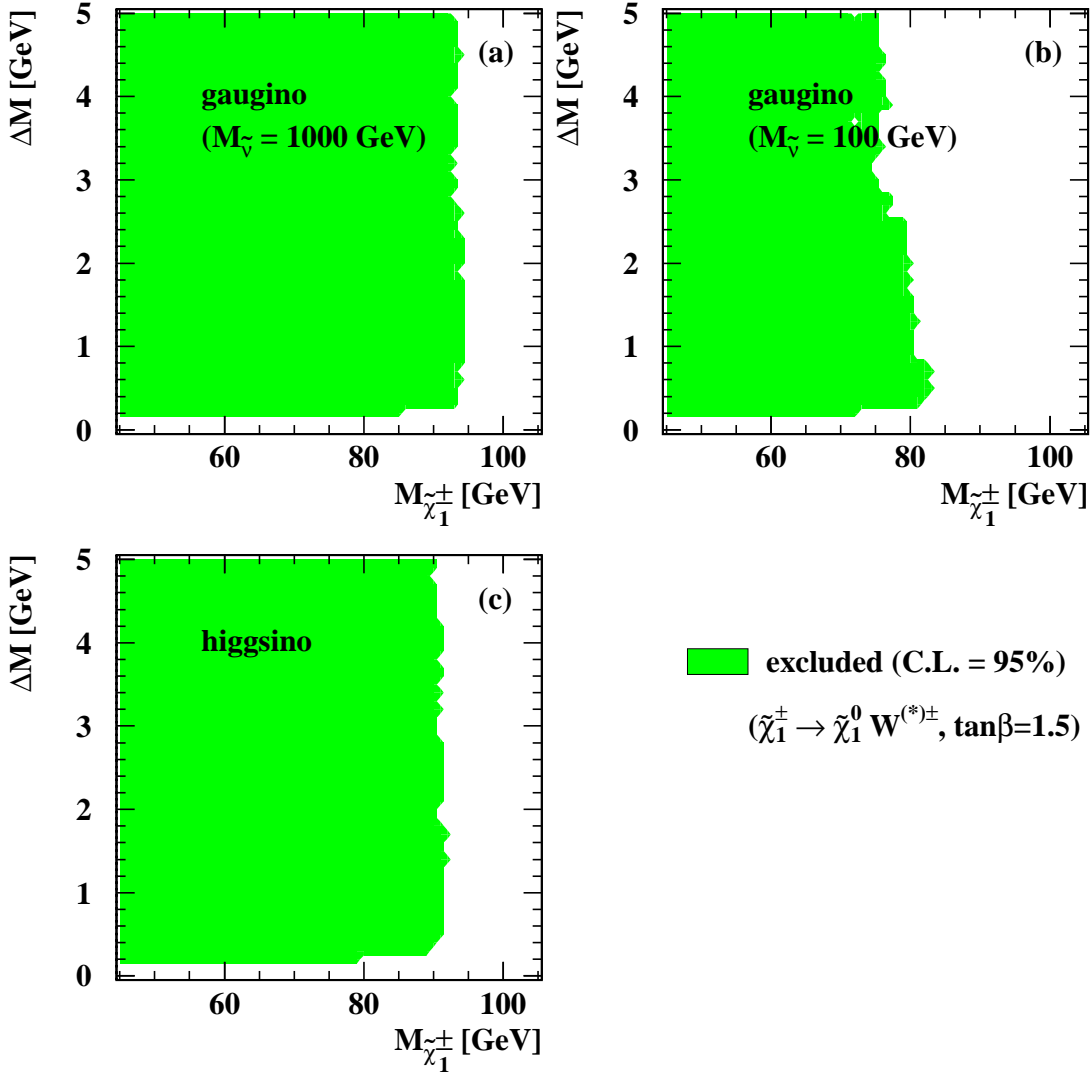}
    \caption{Lower mass limits at the 95\% confidence level for various
      scenarios. These limits are valid for the case of a 100\%
      branching ratio
      $\CHI{\pm}{1}\to\CHI{0}{1}\mathrm{W}^{\pm(\star)}$.  The limits
      shown for $\DM{}{}<0.5\,\GeV$ are only valid with the assumption
      of a zero chargino lifetime. For the gaugino-like scenario with
      a light sneutrino (b) the same signal efficiencies as in the
      case of a heavy sneutrino (a) were assumed.}
    \label{fig:mass-limits}
    \end{center}
\end{figure}

\begin{figure}
    \begin{center}
    \includegraphics[width=.95\textwidth]{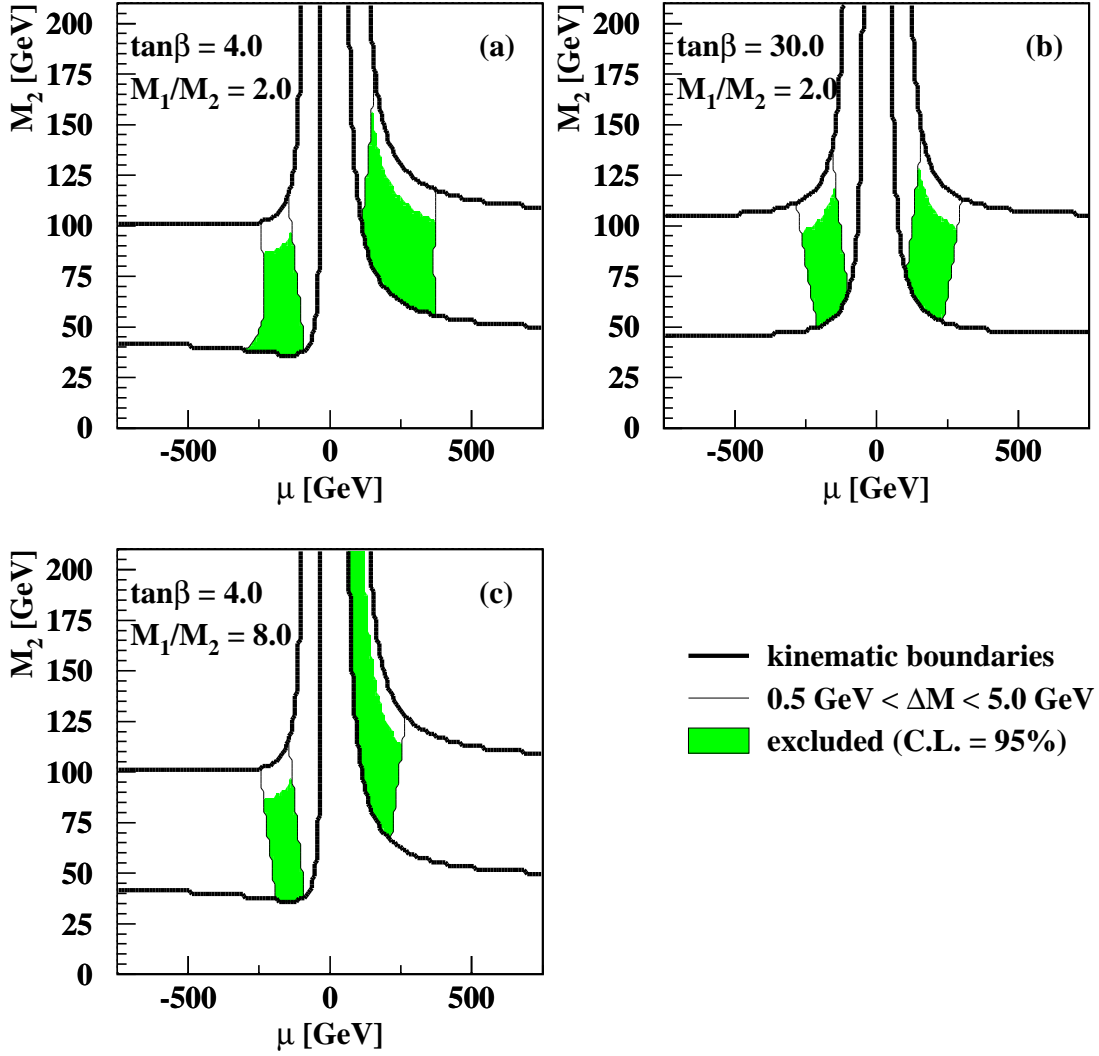}
    \caption{Exclusion regions at 95\% C.L. in the $\mu$-$M_2$ plane within the
      mSUGRA framework for two values of $M_1/M_2$ and two different
      values of $\tan\beta$. The kinematic boundaries
      $(45\,\GeV<\M{\CHI{\pm}{1}}<104\,\GeV)$ are shown by the thick
      line. The thin line indicates the \DM{}{} region considered. The
      shaded areas are the regions excluded by this analysis for
      $\M{\CHI{\pm}{1}}<92\,\GeV$. }
    \label{fig:m2mu-limits}
    \end{center}
\end{figure}

Figure~\ref{fig:xsections} shows the cross-sections used to calculate the mass
exclusion limits for the gaugino- and higgsino-scenarios. Since the coupling of 
the higgsino component of the chargino to the sneutrino is suppressed,
the cross-sections in case of the gaugino-like scenario are
more sensitive to the interfering $t$-channel production and therefore more
sensitive to the sneutrino mass. The chargino mass exclusion regions 
for both scenarios are presented in Figure~\ref{fig:mass-limits}. From 
these results we derive the lower chargino mass limits listed in 
Table~\ref{tab:mass-limits}. The limits are valid for 
$0.5\,\GeV\leqslant\DM{}{}\leqslant5.0\,\GeV$ and zero chargino lifetime
assuming a 100\% branching ratio 
$\CHI{\pm}{1}\to\CHI{0}{1}\mathrm{W}^{\pm(\star)}$.
These mass limits can be directly translated into exclusion regions for 
the above mentioned mSUGRA parameters as shown in 
Figure~\ref{fig:m2mu-limits}. 
 
\begin{table}[htbp]
  \begin{center}
    \begin{tabular}[H]{|c|c|}  \hline
scenario & lower \M{\CHI{\pm}{1}} limit (95\% \CL) \\ \hline
higgsino-like                    & 89 \GeV \\  \hline
gaugino-like                     &         \\
($\M{\tilde{\nu}} = 1000\,\GeV)$ & \RB{92 \GeV} \\  \hline
gaugino-like                     &                 \\
($\M{\tilde{\nu}} = 100\,\GeV)$  & \RB{74 \GeV}  \\ \hline
    \end{tabular}
    \caption{Lower chargino mass limits at the 95\% C.L. within the
             mSUGRA framework. The limits are valid for 
             $0.5\,\GeV\leqslant\DM{}{}\leqslant5.0\,\GeV$ and zero 
             chargino lifetime.}
    \label{tab:mass-limits}
  \end{center}
\end{table}

\subsection{Interpretation within the Anomaly Mediated SUSY Breaking Scenario}
\label{subsec:AMSB}
Alternatives to gravity mediated SUSY breaking scenarios can have
the SUSY breaking not directly communicated from the hidden to the
visible sector, as in the Anomaly Mediated SUSY Breaking Scenario
(AMSB). As already mentioned in Section~\ref{sec:introduction}
anomalies always contribute to soft mass parameters. In AMSB gaugino
masses are generated at one loop and scalar masses at two loops as a
consequence of the super-Weyl anomaly~\cite{Randall:1998uk,
  Giudice:1998xp}. Here we restrict the AMSB models to those models
without any further contributions from other SUSY breaking mechanisms.

In this scenario the chargino is gaugino-like and the $M_1/M_2$ ratio
is $\sim2.8$. AMSB models are described by the following parameters: a
universal scalar mass at the GUT scale, $m_0$; the gravitino mass
$m_{3/2}$; $\tan\beta$ and $\mathrm{sign}\,\mu$.

\begin{figure}
    \begin{center}
    \includegraphics[width=.95\textwidth]{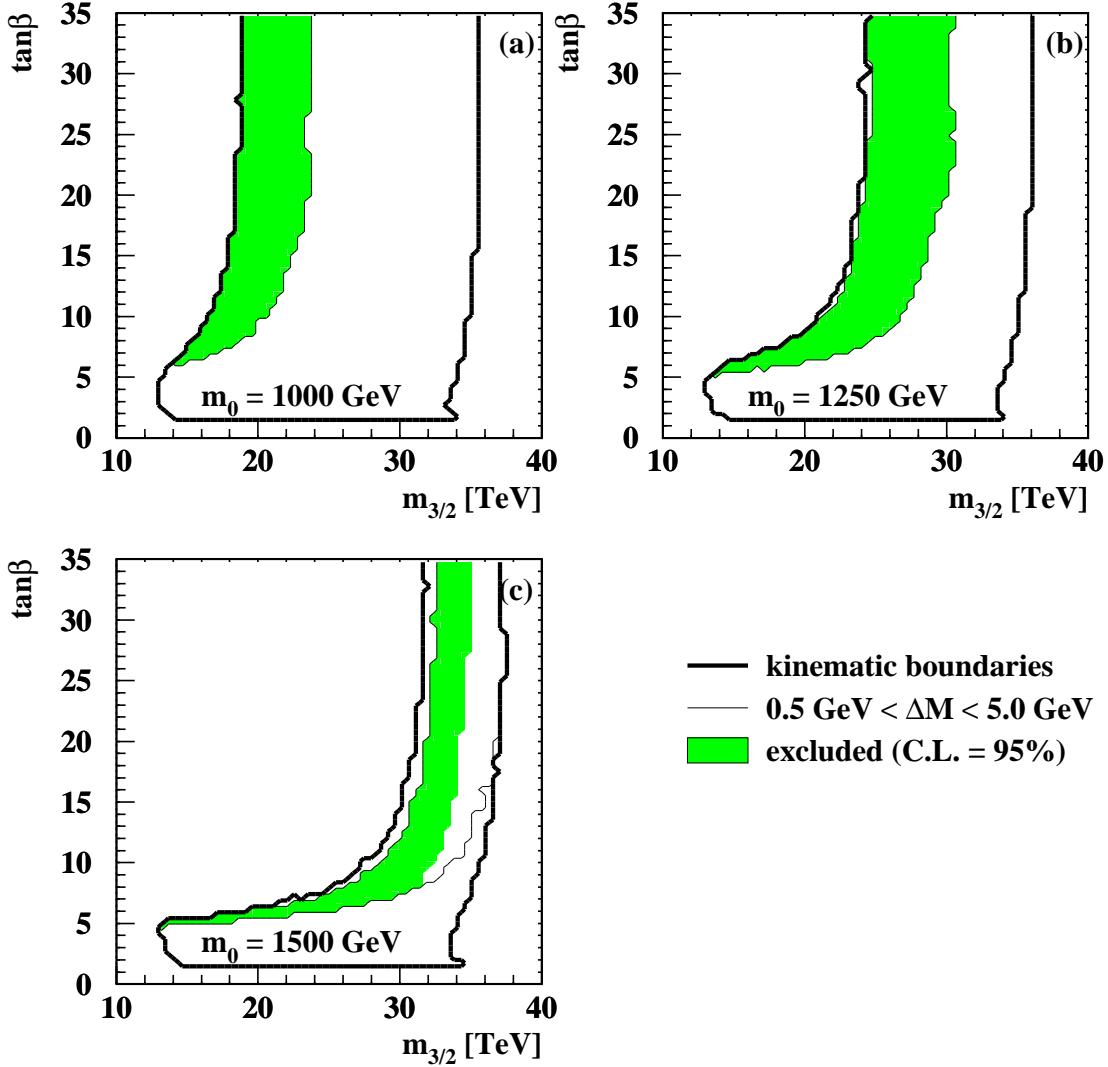}
    \caption{Exclusion regions at 95\% C.L. in the $\tan\beta$-$m_{3/2}$ plane for 
      various values of $m_0$ within the AMSB framework. The kinematic
      boundaries $(45\,\GeV<\M{\CHI{\pm}{1}}<104\,\GeV)$ are shown by
      the thick line. The region below $\tan\beta=1.5$ is theoretically
      inaccessible. The thin line indicates the \DM{}{} region
      considered. The shaded areas are the regions excluded by this
      analysis.}
    \label{fig:amsb-limits}
    \end{center}
\end{figure}

Since SUSYGEN is not able to handle AMSB scenarios, we re-interpreted
our mSUGRA gaugino-like scenario results in order to give exclusion
limits for the AMSB scenario. The ISAJET generator~\cite{Baer:1999sp}
was used to calculate the AMSB spectra which then were mapped to the
corresponding mSUGRA cross-sections.  The corresponding exclusion
regions within the AMSB parameter space are shown in
Figure~\ref{fig:amsb-limits}.

\section{Conclusions}
\label{sec:conclusions}

We have searched for almost mass-degenerate charginos and neutralinos
at centre-of-mass energies between 189 and 209 \GeV with the OPAL
detector at LEP. No significant excess was observed with respect to
the Standard Model background. We derived a lower limit on the
chargino mass of 74~\GeV at the 95\%~\CL for
$0.5\,\GeV\leq\DM{}{}\leq5\,\GeV$ in the case of light sneutrinos
($\M{\tilde{\nu}}>100\,\GeV$) within the mSUGRA framework.
Cross-section and mass limits were translated into mSUGRA and AMSB
parameter exclusion regions.

\section*{Acknowledgements}
We particularly wish to thank the SL Division for the efficient operation
of the LEP accelerator at all energies
 and for their close cooperation with
our experimental group.  In addition to the support staff at our own
institutions we are pleased to acknowledge the  \\
Department of Energy, USA, \\
National Science Foundation, USA, \\
Particle Physics and Astronomy Research Council, UK, \\
Natural Sciences and Engineering Research Council, Canada, \\
Israel Science Foundation, administered by the Israel
Academy of Science and Humanities, \\
Benoziyo Center for High Energy Physics,\\
Japanese Ministry of Education, Culture, Sports, Science and
Technology (MEXT) and a grant under the MEXT International
Science Research Program,\\
Japanese Society for the Promotion of Science (JSPS),\\
German Israeli Bi-national Science Foundation (GIF), \\
Bundesministerium f\"ur Bildung und Forschung, Germany, \\
National Research Council of Canada, \\
Hungarian Foundation for Scientific Research, OTKA T-029328, 
and T-038240,\\
Fund for Scientific Research, Flanders, F.W.O.-Vlaanderen, Belgium.\\


\pagebreak

\pagebreak

\pagebreak

\pagebreak

\pagebreak

\pagebreak

\pagebreak

\pagebreak

\pagebreak

\end{document}